\documentclass{webofc}
\usepackage[varg]{txfonts}   
\usepackage[dvipdfmx]{color}
\begin{document}
\title{Measurements of $p$-$\Xi$, $\Lambda$-$\Lambda$, and $\Xi$-$\Xi$ Correlation in Au+Au collisions at $\sqrt{s_{\rm{NN}}}$ = 200 GeV at RHIC-STAR }
%

\author{\firstname{Moe} \lastname{Isshiki for the STAR Collaboration}\inst{1}\fnsep\thanks{\email{misshiki@rcf.rhic.bnl.gov}}
}

\institute{University of Tsukuba, 1-1-1, Tenno-dai, Tsukuba, Ibaraki, 305-8571, Japan 
          }

\abstract{%

The hyperon-hyperon and hyperon-nucleon interactions are important in search for exotic hadrons such dibaryon as a bound state of two baryons.
$\Lambda$-$\Lambda$ correlation has been measured in Au+Au collisions at $\sqrt{s_{\rm{NN}}}$ = 200 GeV at RHIC-STAR experiment using high statistics data recently taken, which has been found to clearly show the anti-correlation. The first measurements of p-$\Xi$ and $\Xi$-$\Xi$ correlations in Au+Au collisions have been shown. The p-$\Xi$ correlation function shows the strong attractive interaction, which is consistent with HAL QCD calculation. $\Xi$-$\Xi$ correlation has also been measured to be negative (as anti-correlation).

}%
\maketitle
\section{Introduction}
\label{intro}
\par
Hyperon-hyperon (Y-Y) and hyperon-nucleon (Y-N) interactions are important for studying exotic hadrons and to understand the equation of state of neutron stars.
 In case of Y-Y interaction, the existence of H-dibaryon (consisting of six quarks: {\it uuddss}) is predicted by Jaffe in1977 \cite{RefJ}.
Also, the Y-N interaction is interesting because N-$\Xi$ interaction in the spin-singlet and isospin-singlet channel is most attractive according to lattice QCD calculation \cite{RefK}.



Study of $\Lambda$-$\Lambda$ correlation has been performed in STAR \cite{RefS} and ALICE experiments \cite{RefALICE}, but the interaction is still not clear due to large experimental uncertainty and complication of (partly unknown) feed-down contribution. 
In ALICE experiment, the attractive interaction of p-$\Xi$ has been observed in p+p and p+Pb collisions \cite{RefA,RefA2}, which is consistent with a prediction by HAL QCD calculation.
In this analysis, we presents new results of $\Lambda$-$\Lambda$ correlation as well as p-$\Xi$ correlation from high statistics data of Au+Au collisions at $\sqrt{s_{\rm{NN}}}= 200 $ GeV. We also present the first measurement of $\Xi$-$\Xi$ correlation to study a possible bound state or weak attractive interaction of $\Xi$-$\Xi$ system predicted by lattice QCD calculations \cite{RefN, RefD}.



\section{Femtoscopy analysis}
\label{sec-1}
\par

A technique based on Bose-Einstein and Fermi-Dirac correlations has been used in heavy-ion collisions to probe the spatial and temporal extent of particle emitting source. Femtoscopic correlations arise due to quantum statistical and final state effects (strong interaction and Coulomb interaction) at low relative momentum of two particles. The experimental correlation function is defined as
\begin{equation}
C(Q_{\textrm inv})=\frac{A(Q_{\textrm inv})}{B(Q_{\textrm inv})},
\end{equation}
\begin{equation}
Q_{\textrm inv}=\sqrt{{(p_{x1}-p_{x2})}^2+{(p_{y1}-p_{y2})}^2+{(p_{z1}-p_{z2})}^2-{(E_{1}-E_{2})}^2}
\end{equation}
where $A(Q_{inv})$ is the distribution of real pairs from the same events, $B(Q_{\rm inv})$ is the distribution of mixed pairs from two different events as background distribution, and $Q_{\rm inv}$ is relative momentum. Protons and daughter particles of $\Lambda$ and $\Xi$ were identified using Time Projection Chamber (TPC) and Time Of Flight (TOF) detectors, and then $\Lambda$ and $\Xi$ were identified based on decay topology and/or KFParticle package \cite{RefCBM}.
\begin{figure}[t]
 \begin{center}
  \includegraphics[width=110mm]{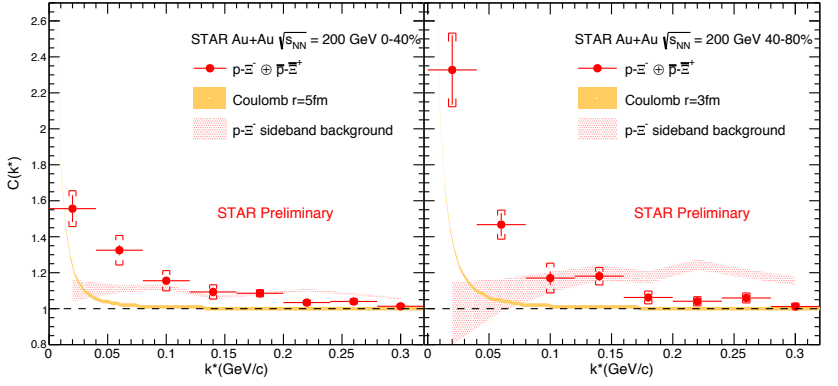}
 \end{center}
 \caption{(Color online) p-$\Xi$ correlation function in 0-40$\%$ and 40-80$\%$ central Au+Au collisions at $\sqrt{s_{NN}}=200$ GeV \cite{KeMi}. The orange bands show the Coulomb strength, varying the source size 5 fm (central collisions) and 3 fm (peripheral collisions). The red hatched bands show the side-band background correlation. }
 \label{fig:one}
\end{figure}

\begin{figure}[htbp]
 \begin{center}
  \includegraphics[width=65mm]{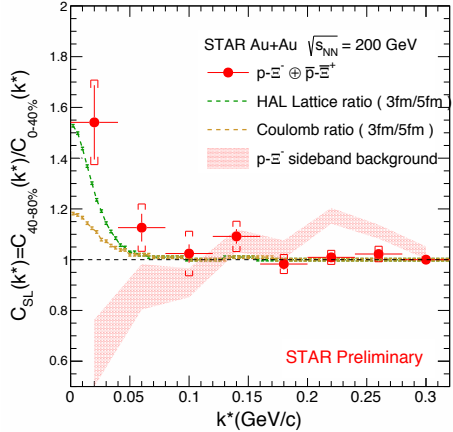}
 \end{center}
 \caption{(Color online) System ratio of the p-$\Xi$ correlation function, in 40-80$\%$ to that in 0-40$\%$ \cite{KeMi}. The green dashed line is the ratio between the source sizes 3~fm to 5~fm calculated by the HAL lattice QCD model. The yellow dashed lines shows the ratio between the source sizes 3~fm to 5~fm estimated for the Coulomb interaction. The red hatched band shows the side-band background correlation.}
 \label{fig:two}
\end{figure}

\section{Results}
\subsection{Measurement of p-$\Xi$ correlation function}
\label{sec-2}
\par
Figure \ref{fig:one} shows the correlation function for p-$\Xi$ pairs combined with their charge conjugate using the data taken in 2010, 2011, and 2014 \cite{KeMi}. The p-$\Xi$ analysis has been done in the pair rest frame. The feed-down effect was corrected using the Theminator2 model \cite{RefT}, but a possible residual correlation is not considered yet. The p-$\Xi$ correlation shows an enhancement compared to the Coulomb interaction. This is a hint for the presence of strong interaction, that can not be described by sideband background. The sideband background could come from other particle pair's correlation outside of invariant mass range. The correlation function in peripheral collisions is stronger than that in central collisions because the system size becomes smaller. Figure \ref{fig:two} shows the ratio of the correlation function in peripheral collision over that in central collision, which is defined as
\begin{equation}
C_{\rm{SL}}(k^{*}) = \frac{C(k^{*})_{40-80\%}}{C(k^{*})_{0-40\%}},
\end{equation}
 where $k^{\ast}$ is half of the relative momentum in the pair rest frame. The ratio $C_{\rm{SL}}(k^{*})$ is more sensitive to the strong interaction because of some partial cancellation of the Coulomb interaction. Below $k^{*}$= 0.1 GeV/c, the signal has been found to be enhanced beyond the Coulomb interaction. The lattice QCD result calculated with Correlation Analysis Tool using the Schrodinger equation (CATS) model \cite{CATS} shows similar behavior to the experimental data which suggests an attractive strong interaction between p and $\Xi$ \cite{RefR}.

\begin{figure}[h]
 \begin{minipage}[]{0.5\hsize}

  \begin{center}
 \mbox{\raisebox{-30mm}{\includegraphics[width=52mm]{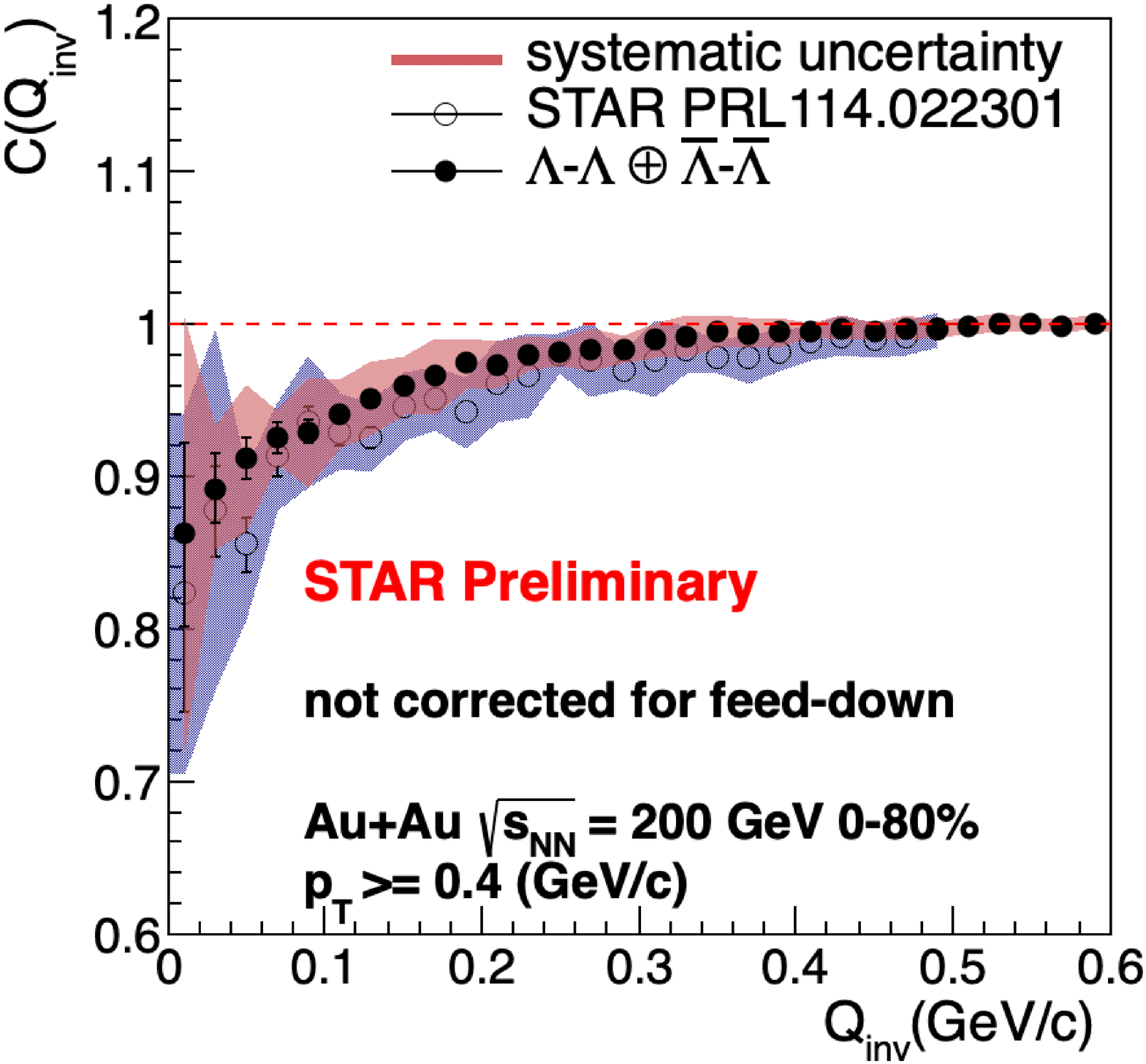}}}
  \end{center}
  \caption{Correlations for $\Lambda$-$\Lambda$ pairs in 0-80$\%$ central Au+Au collisions from the 2011, 2014, and 2016 data sets, requiring $\Lambda$ particles to satisfy $p_{\rm{T}} \geq 0.4$ $\rm{GeV}/c$, |y|<1. The filled symbols are the data of this analysis, and the open symbols are the previous results from STAR. 
 }
  \label{fig:three}
 \end{minipage}
 \hspace{4mm}
 \begin{minipage}[]{0.5\hsize}
  \hfill\vspace{-15pt}
  \begin{center}
  \mbox{\raisebox{5mm}{\includegraphics[width=52mm]{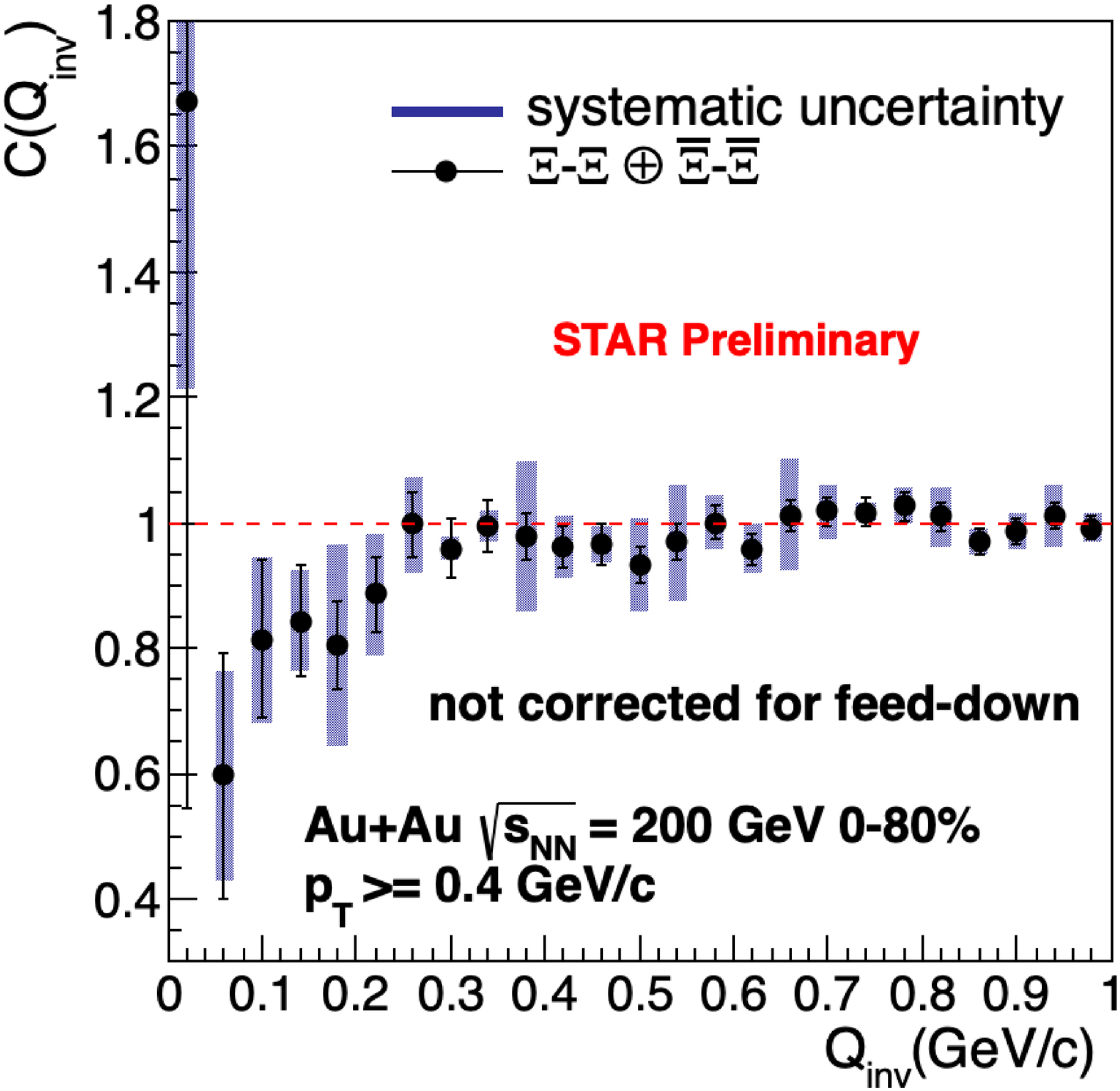}}}
  \end{center}
   \hfill\vspace{-5pt}
  \caption{Correlations for $\Xi$-$\Xi$ pairs in 0-80$\%$ central Au+Au collisions from the 2011, 2014 and 2016 data sets. }
  \label{fig:four}
 \end{minipage}
\end{figure}

\subsection{Measurement of $\Lambda$-$\Lambda$ correlation function}
\par
\label{sec-3}

 Figure \ref{fig:three} is the correlation function for $\Lambda$-$\Lambda$ ($\bar{\Lambda}$-$\bar{\Lambda}$) pairs in Au+Au collisions at $\sqrt{s_{NN}}=200$ GeV using the data taken in 2011, 2014 and 2016.
The statistics used in this analysis is four times larger than that in the previous study \cite{RefS}. This analysis result has not yet been corrected for feed-down. Feed-down means particles originating from weak decays. The feed-down particles can be misidentified as primordial particles.
The new result with better statistical precision is found to be consistent with the previous result within systematic uncertainty. The correlation shows a clear anti-correlation. There is a long tail of residual correlation at high $Q_{\rm{inv}}$, that could be coming from mini-jets and other decays.
The femtoscopic fitting with Lednicky-Lyuboshitz model \cite{RefL} will be applied to extract a specific scattering parameter and it would hopefully clarify the existence of a bound state of $\Lambda$-$\Lambda$ system.

\subsection{Measurement of $\Xi$-$\Xi$ correlation function}
\label{sec-4}

According to NPLQCD calculation \cite{RefN}, the $\Xi$-$\Xi$ is found to be a bound state, while HAL QCD calculation suggests that $\Xi$-$\Xi$ is strongly attractive but not sufficient to form a bound state \cite{RefD}. 
Figure \ref{fig:four} is the first mesurement of $\Xi$-$\Xi$ ($\bar{\Xi}$-$\bar{\Xi}$) correlation in Au+Au collisions at $\sqrt{s_{\rm{NN}}}=200$ GeV. The result is not corrected for feed-down.
The result shows anti-correlation at $Q_{\rm{inv}}$< 0.25 GeV/$c$, which would be given by the combination of quantum statistics, strong interaction and Coulomb interaction. Feed-down and Coulomb effects need to be evaluated for further discussion. More experimental data will be taken in 2023 and 2025 at RHIC-STAR, where much better statistically significant measurement is expected.

\section{summary}
We have reported the first measurements of p-$\Xi$ and $\Xi$-$\Xi$ correlations in Au+Au collisions at $\sqrt{s_{\rm{NN}}}$ = 200 GeV and also revisited $\Lambda$-$\Lambda$ correlations with high statistics data. In the p-$\Xi$ correlation analysis, strong attractive interaction is observed. The correlation function $C(k^{*})$ ratio of peripheral over central collisions, $C_{\rm{SL}}(k^{*})$ is found to be above the Coulomb interaction. This is similar to lattice QCD calculation which suggests an attractive strong interaction between p and $\Xi$. For the $\Lambda$-$\Lambda$ correlation function, new result with high statistics data is presented and consistent with the previous result from STAR, where a clear anti-correlation is observed.
The first mesurement of  the $\Xi$-$\Xi$ correlation seems to show the anti-correlation.

%

\begin{thebibliography}{}
%
%


\bibitem{RefJ}R. L. Jaffe, Phys. Rev. Lett.38, 195 (1977), Phys. Rev. Lett. 38, 617(1977).
\bibitem{RefK}K. Sasaki {\it et al.}, Nuclear Physics A 998,121737(2020).
\bibitem{RefS}L. Adamczyk {\it et al.}, (STAR Collaboration), Phys. Rev. Lett. 114, 022301(2015).
\bibitem{RefALICE}S. Acharya {\it et  al.}, (ALICE Collaboration), Phys. Lett. B 797, 134822(2019) 
\bibitem{RefA}S. Acharya {\it et al.}, (ALICE Collaboration), Phys. Rev. Lett.123, 112002(2019).
\bibitem{RefA2}ALICE Collaboration, Nature 588, 232-238(2020) 
\bibitem{RefN} S. R. Beane {\it et al.},(NPLQCD Collaboration) Phys. Rev. D 85, 054511 (2012).
\bibitem{RefD} T. Doi {\it et al.}, EPJ Web Conf. 175,  05009 (2018).
\bibitem{RefCBM} I. Kisel (CBM Collaboration), J. Phys. Conf. Ser.1070, 012015 (2018).
\bibitem{KeMi} K. Mi et al. STAR, talk presented at APS April Meeting 2021, L13.00007 (2021)
\bibitem{RefT} M. Chojnacki {\it et al.}, arXiv:1102, 0273v1(2011). 
\bibitem{RefM}K. Morita {\it et al.}, Phys. Rev. C 94, 031901(2016).
\bibitem{CATS}D.L.Miaylov {\it et al.}, Eur. Phys. J. C 78: 394 (2018)
\bibitem{RefR} T. Hatsuda {\it et al.}, Nuclear Physics A 967, 856-859 (2017).
\bibitem{RefL} R. Lednicky, V. L. Lyuboshitz, Sov. J. Nucl. Phys. 35, 770(1982).






\end{thebibliography}
%
%

\end{document}